\documentclass[12pt]
{article}
\usepackage{amsmath,amssymb,color,graphicx}
\usepackage{bm}
\usepackage[
colorlinks=true,backref,pagebackref]{hyperref} 
\usepackage{bbm}

\definecolor{dark-green}{rgb}{0,0.7,0}
\definecolor{dark-blue}{rgb}{0,0.2,0.5}
\definecolor{med-blue}{rgb}{0,0.7,1}
\definecolor{mblue}{rgb}{0,0.2,1}
\definecolor{cnc}{rgb}{0.8,0,0}
\definecolor{light-red}{rgb}{1,0.8,0.8}
\definecolor{dark-yellow}{rgb}{1,0.8,0}
\definecolor{light-blue}{rgb}{0.8,0.9,1}
\definecolor{verylight-blue}{rgb}{0.93,0.95,1}
\definecolor{light-yellow}{rgb}{1,0.9,0.8}
\definecolor{grey}{gray}{0.88}
\definecolor{mygray}{rgb}{0.5,0.5,0.5}

\def\a{\alpha}

\def\ve{\varepsilon}

\begin{document}

\title{\bf Physical dimensions/units and universal constants: their
  invariance in special and general relativity\footnote{In ``The
    Revised SI: Fundamental Constants, Basic Physics and Units,''
    Proceedings of the 670th WE-Heraeus-Seminar in Bad Honnef, 13 to
    18 May 2018, K.~Blaum, D.~Budker, A.~Surzhykov, and J.~H.~Ullrich,
    editors, special issue of the Annalen der Physik (Berlin),
    Wiley-VCH, to be published 2019. Our article can be found under
    https://doi.org/10.1002/andp.201800407 .}}

\author{Friedrich W.\ Hehl$^1$ and Claus L\"ammerzahl$^2$\\
  $^1$Institute for Theoretical Physics, University of Cologne\\ 50923
  K\"oln, Germany, \begin{footnotesize}email: hehl@thp.uni-koeln.de
  \end{footnotesize}\\
  $^2$Center of Applied Space Technology and Microgravity (ZARM),
  \\University of Bremen\\28359 Bremen, Germany, \begin{footnotesize}
    email: laemmerzahl@zarm.uni-bremen.de\end{footnotesize}}

\date{\begin{footnotesize} 01 Feb 2019, {\it file
      BadHonnef\_FundConsts2018\_19.tex}\end{footnotesize}}
\maketitle
\begin{abstract}
  The theory of physical dimensions and units in physics is
  outlined. This includes a discussion of the universal applicability
  and superiority of quantity equations. The International System of
  Units (SI) is one example thereof. By analyzing mechanics and
  electrodynamics, we are naturally led, besides the dimensions of
  length and time, to the fundamental units of action $\mathfrak h$,
  electric charge $q$, and magnetic flux $\phi$. We have
  $q\times \phi=\text{action}$ and $q/\phi=1/\text{resistance}$.
  These results of {{\it classical physics}} suggests to look into the
  corresponding quantum aspects of $q$ and $\phi$ {(and also of
    $\mathfrak h$)}: The electric charge occurs exclusively in
  elementary charges $e$, whereas the magnetic flux can have any
  value; in specific situations, however, in superconductors of type
  II at very low temperatures, $\phi$ appears quantized in the form of
  fluxons (Abrikosov vortices). {And $\mathfrak{h}$ leads, of course,
    to the Planck quantum $h$.} Thus, we are directed to
  superconductivity and, because of the resistance, to the quantum
  Hall effect. In this way, the Josephson and the quantum Hall effects
  come into focus {quite naturally}. One goal is to determine the
  behavior of the fundamental constants in special and in general
  relativity, that is, if gravity is thought to be switched off versus
  the case in the gravitational field.
\end{abstract}

\maketitle


{\hypersetup{linkcolor=black}
\tableofcontents}
\medskip

\noindent{\bf Keywords:} Physical dimensions, units, SI, universal constants,
  relativistic invariance, speed of light, Josephson effect, quantum
  Hall effect.

\section{Introduction and summary}\label{Sec.1}
\noindent String theory at its best (G.\ Veneziano, 2002, see
\cite{dov}): \vspace{2pt}

{\it ``... it looks unnecessary (and even ``silly'' according to the
  present understanding of physical phenomena) to introduce a separate
  unit for temperature, for electric current and resistance,
  etc...''}\\

Next year an essential reform of the International System of Units
(SI) will be enacted: In particular the {\it kilogram} (kg) will no
longer be realized by a physical artifact, the international prototype
of the kilogram kept in S\`evres near Paris, see the talks of Quinn
\cite{Quinn:2018,Quinn:2018a} and Ullrich \cite{Liebisch:2018} and
also \cite{Quinn:2013,Wikipedia:2018}. Rather the mass will be linked
via the Kibble (or watt) balance to electromagnetic units to be
measured by the Josephson effect (JE) and the Quantum Hall Effect
(QHE), see von Klitzing and Weis
\cite{KlitzingReview:2017,WeisKlitzing:2011} and G\"obel and Siegner
\cite{Goebel:2015}. The JE yields the Josephson
constant\footnote{Recall that P = peta = $10^{15}$, and note that Hz/V
  corresponds dimensionally to
  $1/{\rm magnetic\, flux}\stackrel{{\rm SI}}{=}1/{{{\rm weber}}}$.}
$K_{\rm J}=2e/h\stackrel{{\rm SI}}{\approx}0.483\,{\rm PHz/V}$ and QHE
the von Klitzing constant
$R_{{\rm K}}={h/e^2}\stackrel{{\rm SI}}{\approx}25.813\;{\rm
  k}\Omega$.  Here $h$ is the Planck constant,\footnote{Since ``it
  cannot be logically excluded that in different realms of physics are
  in fact described by distinct quantization constants,...'' Fischbach
  et al.\ \cite{Fischbach:1991eg} checked it experimentally with high
  accuracy that the Planck constant is universally valid.}
$\hbar:=h/(2\pi)$, and $e$ the elementary electric charge. {As it is
  clear from the definitions of $K_{\rm J}$ and $R_{\rm K}$, these two
  constants alone can directly determine the elementary charge $e$ and
  the Planck constant $h$, without the intervention of any other
  constant.}

Besides $h$ and $e$, the {\it speed of light} $c$ belongs to the
fundamental constants of nature. Derived from $h$, $e$, and $c$, we
find Sommerfeld's fine structure constant
$\alpha\stackrel{{\rm SI}}{:=}e^2/(2\varepsilon_0 \,c \,h)$, with
$\varepsilon_0$ as the electric constant (``permittivity of free
space''). The dimensionless constant $\alpha\approx 1/137$, the
coupling constant of the electromagnetic field in quantum
electrodynamics, can alternatively be expressed in terms of the
quantum measure $R_{\rm K}$ as
$\alpha\stackrel{{\rm SI}}{=}\Omega_0/(2R_{\rm K})$, with
$\Omega_0 \stackrel{{\rm SI}}{\approx}377\,\Omega$ as the vacuum
impedance. Thus, with the help of the von Klitzing constant
$R_{\rm K}$, the fine structure constant does not depend on the speed
of light $c$ any longer.\footnote{Usually, a possible time dependence
  of $\alpha=\alpha(t)$, see Uzan \cite{Uzan:2002vq}, is linked to the
  question whether $c$ may be time dependent, too. This is problematic
  anyway since $c$ carries a dimension. But with
  $\a=\Omega_0/(2R_{\rm K})$, we recognize that in the new SI the
  speed of light $c$ drops out in $\a$ altogether.}  {A popular
  exposition of the different fundamental constants of nature has been
  given by Barrow \cite{Barrow}.}

Conventionally, the universal constants $e$, $h$, and $c$---and thus
also $K_{{\rm J}}$ and $R_{{\rm K}}$---are tacitly assumed to be
scalars, that is, they are totally independent of the coordinate
systems or reference frames chosen for the description of the
corresponding experimental arrangements. For $h$, e.g., this can be
read off from the Einstein and the deBroglie relations,
$E=\hbar \omega,\; p_i=\hbar k_i$; since the momentum 4-covector
$p_\mu=(E,p_i)$ is related to the wave wave 4-covector
$k_\mu=(\omega,k_i)$ according to $p_\mu=\hbar k_\mu$, see Rindler
\cite{Rindler:2001}; here $\mu=0,1,2,3$ and $i=1,2,3$. Since $p_\mu$
and $k_\mu$ are 4d covectors in special relativity (SR) as well as in
general relativity (GR), $\hbar$ and $h$ have to be GR-scalars.

For the speed of light $c$, there arises a problem. Since 1919, when
the deflection of light by the Sun was observationally established, it
was clear, however, that the speed of light in a gravitational field
is different from its vacuum value $c$. After all, the gravitational
field acts like a refracting medium with an index of refraction
$n\ne 1$. Still, in SI, the speed of light is assumed to be a
universal constant. As we will discuss in our article, we should
denote the universal constant in SI by $c_0$. Only if gravity can be
neglected, we have $c_0=c$. In other words, in GR, Einstein's theory
of gravity \cite{Meaning}, $c_0$ is a GR-scalar, but $c$, the speed of
light, is not; it is only a scalar in the context of SR, the theory of
spacetime if gravitation can be neglected. Thus, $c$, the speed of
light, is a SR-scalar only---in contrast to what is stated in
SI. After all, in SI, $c$ as speed of light is assumed to be a
universal constant.

This discussion of the universal nature of $c_0$ was foreshadowed by a
ground breaking article of Fleischmann \cite{Fleischmann:1971}. He
pointed out that in physics there are on one side 4-dimensional (4d)
laws that do {\it not} contain the metric $g_{ik}$ and are covariant
under general coordinate transformations (diffeomorphism covariant),
on the other side those 4d laws in which the metric $g_{ik}$ is
involved. To the former belong the Maxwell equations, to the latter
their constitutive relations and the Einstein field equation of
gravity, see Post \cite{Post:1962}.

As we will find out, there exist only a few GR-scalars, namely action
$W$, electric charge $q$, magnetic flux $\phi$, entropy $S$, as well
as products and quotients thereof. Interestingly enough, in nature,
the values of all of those true 4d scalars can be changed via
quanta. Note, Hamiltonians, energies, and energy densities, e.g., are
{\it not} 4d scalars. We know only one example of a SR-scalar, namely
the speed of light $c$. The reason for this exceptional status of $c$
seems to be that $c$ is defined in terms of the pre-relativistic
notions of {\it time} and {\it space}.

In order to base our analysis on an up-to-date view of the theory of
dimensions and units, we start our discussion with those notions in
Sec.2. In particular, we will concentrate on so-called quantity
equations, which are valid in all systems of units, whereas the
numerical equations, often used in quantum field theory, are only valid
in one specific system of units. The SI is based on the notion of a
{\it physical quantity} and the corresponding {\it quantity
  equations.} This quantity calculus will be exclusively used by us.

In Sec.3 the basic physical dimensions of classical Newtonian
mechanics are introduced and in Sec.4, by going over to the
Lagrange-Hamilton formalism, the action is derived as a new basic
element. Needless to say that the notion of an action emerges in
classical physics, well before quantum theory was discovered. 

With the emergence of magnetism and electricity, totally new physical
phenomena were discovered. In Sec.5, following basically Giorgi, the
electric charge $q$ is introduced as a new fundamental physical
quantity. In Sec.6, the Maxwell equations are formulated in 3- and in
4-dimensional formalisms. Absolute and relative physical dimensions
are introduced and, besides the electric charge, the magnetic flux
$\phi$ identified as a second fundamental electromagnetic quantity.
In Sec.7 the constitutive relations of electrodynamics are formulated
for local and linear matter. A constitutive tensor of fourth rank with
the dimension of an admittance is found. In Sec.8, following our
discussion on dimensions, quantum aspects of electric charge and
magnetic flux are displayed, which lead, in Sec.9, in a direct way to
the {\it Josephson} and the {\it von Klitzing} constants. These
constants, together with the Kibble balance, roughly sketched in in
Sec.9, lead directly to the new SI. Finally, in Sec.10, we discuss
critically the status of the speed of light as SR-scalar and, in
Sec.11, collect our ideas on the possible further development of the
new SI.

Summing up, it is near at hand to reconsider the relativistic
invariance with and without gravity of $h$, $e$, and $c$, but also of
$R_{\rm K}$ and $K_{\rm J}$. With the exception of $c$, all these
quantities are GR-scalars. The speed of light $c$, however, is only a
SR-scalar, a fact apparently not appreciated by the SI authorities.


\section{Physical quantity, physical dimensions, quantity equations}\label{Sec.2}

A {\it physical quantity} can always be represented by a {\it number}
or, in the case of spinors, vectors, tensors, by numbers, and by a
{\it physical dimension.} One can read the physical dimension as a
code of how to measure this quantity. The number/s depend on what
system of units is chosen. A unit (like `V' or `kg') should be clearly
distinguished from a physical dimension (here `voltage' or `mass',
respectively).

Dimensional analysis can lead to a better understanding of the {\it
  structure} of a physical theory, see Bridgman \cite{Bridgman:1931},
Wallot \cite{Wallot:1953}, and Stille \cite{Stille:1961}. We have
\begin{eqnarray}\text{Physical quantity} 
  &=&\text{  some numerical value $\times$ unit}\nonumber \\
  Q&=&\{Q\}\times [Q].
\end{eqnarray} 
Example: $T=23\; {\rm h} =23\times 60\; {\rm min}=...\,$. As a consequence, we find
the inverse proportionality rule:
\begin{equation}Q=\{Q\}'\times [Q]'=\{Q\}''\times [Q]''\, \text{  or }\, 
\{Q\}'/\{Q\}''= [Q]''/ [Q]'\,.
\end{equation}
A physical quantity is invariant with respect to the choice of
units. 

The set of all possible units may be called the {\it dimension} of a
quantity, here $[T]=$ time. It is a {\it qualitative} aspect of a
physical quantity. To repeat it: A physical dimension encodes the
knowledge of how to set up an experiment to {\it measure} the
quantity.

{Physical quantities, which are valid for all units, build up
  relations or laws in physics. We call them quantity equations and
  the corresponding calculus is called {\it quantity calculus.} The
  algebraic structure of quantity calculus has been investigated, for
  instance, by Fleischmann
  \cite{Fleischmann:1951,Fleischmann:1959/60}, G\"ortler
  \cite{Gortler}, Emerson \cite{Emerson}, Janyska, Modugno, Vitolo
  \cite{Janyska:2009}, and Kitano \cite{Kitano:2013aia}. For a
  pedagogical introduction we recommend Robinett \cite{Robinett},
  e.g..} A critical analysis of the notion of the physical dimension
shows that a mathematically more rigorous definition is desirable,
see, for instance, Krystek \cite{Krystek:2015}. He opts, inter alia,
for the introduction of a `dimension number' within SI.

The {\it SI} (International System of Units) is based on the quantity
calculus.  Its history has been recorded by de Boer \cite{deBoer}.

{\it Numerical value equations}, used mainly by particle physicists
($c=1,\hbar=1,\kappa_{\rm grav}=1$), are only valid for certain units;
usually insight is then lost into the physical structure and the
interpretation of the corresponding theory.


\section{Physical dimensions in classical mechanics}\label{Sec.3}

The fundamental notions in physics are set up in the context of
Newtonian mechanics. Let us start with

\noindent$\bullet\,${\bf length $\ell$} (in SI $\rightarrow$ m), area $A$,
and volume $V$, namely the fundamental dimension and those derived
therefrom. Dimension of $\ell$ is $[\ell]$.  Compare the length of a
segment 1 with that of a segment 2:
\begin{equation}\label{length1}\{L_2\}_{\it cm}/\{L_1\}_{\it cm}=
\{L_2\}_{\it inch}/\{L_1\}_{\it inch}\,.
\end{equation}
The ratio of two lengths is invariant under the change of units. Length
is additive, so are the derived dimensions $[$area$]=[$length$]^2$,
$[$volume$]=[$length$]^3$. Length is used here in the sense of a
segment in flat {\it affine geometry}, before the distance concept
(metric) is defined, see Choquet \cite{Choquet:1970}. Thus, the affine
length in dimensional analysis is a {\it premetric} concept. Next is

\noindent$\bullet\,${\bf time $t$} (in SI $\rightarrow$ s).
{Because of its unrivaled accuracy, frequency measurements
  are decisive in many aspects of modern metrology, see Flowers
  \cite{Flowers}. In 1968, one was eventually led to the redefinition
  of the second from one based on the rotation of Earth to an atomic
  one: ``The second is the duration of 9 192 631 770 periods of the
  radiation corresponding to the transition between the {\it two
    hyperfine levels} of the ground state {\it of the caesium 133
    atom,}'' see also G\"obel and Siegner \cite[p.11]{Goebel:2015}. The
  length is then defined via a fixed value of $c$ in terms of this new
  second. The definition of the second in terms of a frequency is then
  the fundamental dimension in SI; the meter is a derived concept. The
  velocity $\mathbf{v}$, with $[v]=[\ell]/[t]$, and the acceleration
  $\mathbf{a}$, with with $[a]=[\ell]/[t]^2$, are both derived
  dimensions.}
 
Newton introduced for the {\it quantity of matter} (`quantitas
materiae') the

\noindent$\bullet\,${\bf mass $m$} (in SI $\rightarrow$ kg) and moreover the 

\noindent$\bullet\,${\bf force $f$} (in SI $\rightarrow$
N = kg\,m/s$^2$), measured by means of a a spring scale, for example.

On the basis of Newton's equation of motion, we can interrelate length
$\ell$ , time $t$, mass $m$, and force $f$. Independent {dimensions}
are, for instance, $(\ell,t,m)$, as one assumed in SI.

\section{Lagrange-Hamilton formalism: action $\mathfrak h$ as a new
  physical dimension}

Nowadays in classical mechanics we should use {\it time, length,} and

\noindent$\bullet\,${\bf action} $\mathfrak h$ (in SI $\rightarrow$
J\,s) 
as fundamental dimensions (and in classical electrodynamics,
additionally, electric charge and magnetic flux, see below). Already
Post \cite{Post:1962} argued strongly and convincingly that one should
replace already in classical mechanics the dimension of mass $m$ by the
dimension of action $\frak{h}$. The action is a scalar in special
relativity (SR) and in general relativity (GR) alike; it can be cut
into scalar `portions.'  Time and length are of a different character,
see, e.g., Tonti \cite{Tonti:2013}.

Accordingly, the scalar action function with the dimension of action
$\frak h$ surfaces as a new type of dimension. Thus, following Post
\cite{Post:1962}, we can opt in classical mechanics alternatively for
$(l,t,\frak{h})$ as a basis set of dimensions. $\frak{h}$ is
relativistically invariant also in special relativity (SR) and general
relativity (GR). Thus, this set is more adapted to relativistic
conditions (high velocities etc.) than the set $(\ell,t,m)$.

The mass was experimentally determined by Lavoisier (1789) to be
conserved, however, in SR and GR alike, this rule is broken, see
Jammer \cite{Jammer:1961}. The explosion in Alamogordo (1945) by
nuclear fission was a more than clear demonstration of the energy mass
equivalence and thus of the non-conservation of mass. Note (in the
parentheses we use SI units),\footnote{One may suggest `{planck}' as
  a new SI unit for action.}
\begin{eqnarray}
  &&  \text{action }\frak{h} \nonumber\\
  &=&\mbox{energy}\times\mbox{time}\quad ({\rm J\,s=Pl}=
      \text{planck})\nonumber\\
  &=&\mbox{momentum}\times
      \mbox{length}\quad
      ({\rm kg\times{m}\,{s}}^{-1}\times {\rm m=N\,m\,s=J\,s})
      \nonumber\\
  &=&\mbox{angular momentum}
      \quad  ({\rm kg\times{m}\,{s}}^{-1}\times {\rm m=N\,m\,s=J\,s})\;\nonumber\\
  &=& \mbox{el.\ charge}\times\mbox{magn.\ flux}\quad 
      ({\rm C\times W\hspace{-1pt}b=A\,s\times V\,s=J\,s})\,.
\end{eqnarray}
Incidentally, Bohr was inspired to set up his model for the atom when
he recognized that the dimensions of the action and the angular
momentum are the same, a coincidence that, as far as we are aware, is
still {unexplained}. Hence the equations listed above can
be very helpful at times for winning additional insight in the inner
working of nature. {A similar problematic case we have,
  e.g., with heat capacity and entropy: These quantities are of a
  different kind, still, they carry the same physical dimension.}

\section{Metrology in electromagnetism: electric charge 
$q$ as a new physical dimension}

The discovery of new phenomena---we will turn now our attention to
magnetism and electricity---requires an extension of the theory of
dimensions developed so far. But it took a long time until that became
appreciated.

Let us first drop a few names of main players who were closely
connected with the development of a dimensional analysis in
electromagnetism:
Gauss... Weber... Maxwell... Heaviside... Helmholtz... Hertz... Planck...
Giorgi... Wallot... $\rightarrow$ SI (International System of
Units). For a full-fledged historical perspective, compare Whittaker
\cite{Whittaker:1953:1} and Quinn \cite{Quinn:2013}.

Gauss and Weber (around 1840) recognized the need for precise
measurements in electromagnetism and performed some of these. In
particular, in the Weber-Kohl\-rausch experiment (1856) the speed of
light was measured by sheer electric and magnetic measurements
alone. This result was used by Maxwell in setting up his theory of the
electromagnetic field (1865). Following earlier work of Fourier within
the theory of heat conduction,\footnote{This was pointed out to us by
  M.~P.~Krystek (Berlin) in a private communication.} Maxwell
recognized the need for a {\it physical quantity} as part of the
formulas in electrodynamics. Hence Maxwell may be regarded as the
central figure of the theory of physical dimensions in
electrodynamics.  However, in the 19th century (see also Planck
\cite{Planck:1899}), electromagnetic units were supposed to be reduced
to {\it mechanical} measurements (Gauss units). This prejudice,
originating from classical mechanics, propagated well into the 21
century up to `modern' textbooks of electromagnetism (Jackson).

However, already Giorgi (1901) had cut the Gordian knot: He postulated
the need for an independent electrical dimension, see Frezza et
al.\cite{Frezza}. As his first choice, Giorgi mentioned the electric
resistance ($\Omega$). Thus, he intuitively selected a GR-scalar as
new fundamental electric quantity. The von Klitzing constant
$R_{\rm K}$, see below, has exactly the same physical dimension!
Later, however, the

\noindent$\bullet\,${\bf electric charge} $q$ (in SI $\rightarrow$
${\rm C = A\hspace{1pt}s}$) and eventually the {\it electric current} $j$
were adopted by the international committees. Only in 2019, the new
SI, {via the von Klitzing constant,} will go back to a
resistance as fundamental.

{There occurred a curious historical intermezzo in the
  measurements of the electric charge and the electric current. Until
  1947, the coulomb was defined as fundamental unit of electric charge
  via the {\it mass} of the silver disposed electrolyticly by a
  constant current flowing through a specified aqueous solution of
  silver nitrate AgNO$_3$, see Frezza \cite{Frezza}. This definition
  is consistent with our modern understanding of charge as an
  ``extensity'' (how much?). The electric current in ampere A was
  then a derived unit defined as A=C/s.

  In the intermediate time, till 2019, for reasons of practicality,
  the A was taken as fundamental unit defined via the {\it force}
  between two current carrying wires. Thus, by the definition of
  charge as C=As, the charge was reduced to an ``intensity'' (how
  strong?)---a step backwards to the understanding of the charge at
  the beginning of the 19th century (see the old-fashioned and
  outdated Gauss system of units). Only in the new SI, starting in
  2019, electric charge becomes again interpreted as an {\it
    extensity.} And this is exactly how it should be.}

Nowadays, we can count single electrons with nano-technical tools,
verifying one of Giorgi's hypotheses that a `portion' of an electric
charge can also be used as a new fundamental concept. A basic set for
dimensional analysis is now, for example, $(l,t,\frak{h},q)$, as
presented in the book of Post \cite{Post:1962}, see also
\cite{Hehl:2004jn,Hehl:2016glb,Itin:2016nxk}.

In particular Wallot \cite{Wallot:1953} in the 1930s developed the
Maxwellian idea of {\it quantity equations}, cf.\ also Schouten
\cite{Schouten:1989}. All of this led to the modern SI
($\sim$1955). { As already mentioned, a description of the
  history of quantity equations has been given by de Boer
  \cite{deBoer}.}

\section{Dimensional analysis of the Maxwell equations:
  magnetic flux
  $\phi$ as a further new physical dimension}

Because of the importance of electromagnetism, we will analyze the
field equations of electrodynamics, the Maxwell equations, a bit
closer. This yields at the same time an appropriate dimensional
analysis, see Schouten \cite{Schouten:1989} and Post \cite{Post:1962},
but also Puntigam et al.\ \cite{Puntigam:1996wh} and Obukhov et al.\
\cite{Birkbook}. We will denote the electric excitation (``electric
displacement'') by $\bm{\mathcal{D}}$ and the magnetic excitation
(``magnetic field'') by $\bm{\mathcal{H}}$, furthermore the electric
and the magnetic field strengths by $\bm E$ and $\bm B$,
respectively. We can visualize these fields by pictograms
\cite{Schouten:1989}, see Figure 1.

\begin{figure}
\includegraphics[width=9truecm]{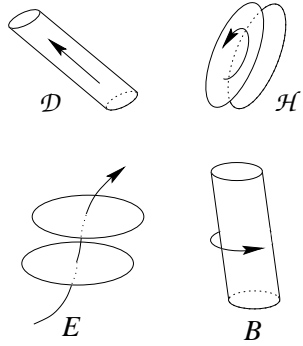}
\caption{\label{Fig.1}Faraday--Schouten pictograms of the
    electromagnetic field in 3-dimensional space. The
      images of the two 1-forms $E$ and $\mathfrak{H}$ are represented
      by two neighboring planes, those of the two 2-forms
      $\mathfrak{D}$ and $B$ by flux tubes. The nearer the planes, the
      stronger is the 1-form; the thinner the tubes, the stronger is
      the flux. The twisted forms, the excitations $\mathfrak{D}$ and
      $\mathfrak{H}$, are source variables and carry an outer, the
      untwisted ones, the field strengths $E$ and $B$, are
      configuration variables and carry an inner orientation. For
      details compare \cite{Schouten:1989,Birkbook} and
      \cite[p.116]{Tonti:2013}.}
\end{figure}

\noindent As sources we have the electric charge density $\rho$ and
the electric current $\bm{j}$. With
($\bm{\partial}\times\rightarrow \text{curl},\bm{\partial}\bm{\cdot}
\rightarrow \text{div}$)
and a dot over a field as partial derivative with respect to time, the
Maxwell equations read as follows (compare this with the Tonti diagram
\cite[p.312]{Tonti:2013}):
\begin{equation}\label{13Maxwell}\text{
\noindent
\begin{tabular}{|l|c|}
  \hline
  Physics law   & Math. expression  \\ \hline\hline
  Amp\`ere-Maxwell law&  
  $\!\bm{\partial}\times \bm{\mathcal{H}} - \bm{\dot\mathcal{D}} =
  \bm{j}$\\ 
  \hline Coulomb-Gauss law&\hspace{25pt}
  $\bm{\partial}\cdot\bm{\mathcal{D}} = \rho$ \\ \hline
  Faraday induction law& $\bm{\partial}\times \bm{E} + \bm{\dot{B}} =
  0$ \\ \hline
  cons.\ of magnetic flux&\hspace{25pt} $\bm{\partial}\cdot\bm{B}  =
  0$  \\ \hline
\end{tabular}
}\end{equation}

\noindent This is the premetric form of the Maxwell equations, that
is, they are totally independent of the metric and are valid in this
form in SR and in GR likewise. The metric $g_{\mu\nu}$ enters only the
constitutive relations linking the excitations to the field
strengths. In {\it vacuum,} we have
$(\mathfrak{g}:=\sqrt{-{\det g_{\mu\nu}}})$:
\begin{equation}\label{vacuum}\text{
\noindent
\begin{tabular}{|l|c|}
 \hline
  permittivity of vac.\ $\varepsilon_0$ & $\bm{\mathcal{D}}
      =\hspace{5pt}\varepsilon_0\,
  \hspace{5pt}\mathfrak{g}\,\hspace{5pt}\bm{E}$\\ \hline
  permeability of vac.\ $\mu_0$ & \hspace{5pt}$\bm{\mathcal{H}}=\mu_0^{-1}\,
                               \mathfrak{g}^{-1}\bm{B}$ \\ \hline
\end{tabular}
}\end{equation}
{The scalar density $\mathfrak{g}$ is necessary here
  in order to consistently link the source to the configuration variables.}
In a 4d calculus, we can collect the inhomogeneous and the homogeneous
Maxwell equations into one 4d inhomogeneous and one 4d homogeneous
Maxwell equation, respectively. For conciseness, we take here the
calculus of exterior differential forms. The 4d twisted excitation
2-form $G=(\bm{\mathcal{D}},\bm{\mathcal{H}})$ and the 4d field strength
2-form $F=(E,B)$ are defined as follows (for details see
\cite{Birkbook}, for example):
\begin{equation}\label{GF4d}
G=\bm{\mathcal{D}}-\bm{\mathcal{H}}\wedge dt\,,\quad F=B+E\wedge dt\,.
\end{equation}
Here $\wedge$ denotes the exterior product and $t$ the time.
Then, with the current 3-form $J=\rho-j\wedge dt$, the two Maxwell
equations in 4d language read
\begin{equation}\label{Max4d}
dG=J\,,\quad dF =0\,.
\end{equation}
From an axiomatic point of view, see \cite{Birkbook}, one may consider
electric charge and magnetic flux conservations $dJ=0$ and $dF=0$,
respectively, as fundamental axioms. The inhomogeneous Maxwell
equation, $J=dG$, can then be considered as an ansatz for solving
$dJ=0$; additionally, the excitation $G$ turns out to be a measurable
quantity---and this makes $G$ as something more than just a
potential. This view on electrodynamics makes it comprehensible why
electric charge $q$ and magnetic flux $\phi$ are the fundamental
dimensions in electrodynamics.

Let us now introduce the notion of {\it absolute and relative
  dimensions:} Absolute dimensions are assigned to a 4d physical
quantity, relative dimensions are those of the components of this
quantity with respect to a local tetrad $\mathbf{\vartheta}^\alpha$
(coframe), with the ``legs''
$(\vartheta^0,\vartheta^1, \vartheta^2,\vartheta^3)$.
$[\vartheta^0]= $ time, $[\vartheta^a]=$ length, for $a=1,2,3$.

Let us first turn to mechanics: The 4-momentum of a particle with mass
$m$ and velocity $\mathbf{v}$ reads
$\mathbf{p}=p_\alpha
\vartheta^\alpha=p_0\vartheta^0+p_1\vartheta^1+p_2\vartheta^2+p_3\vartheta^3$.
Assume for $\mathbf{p}$ the absolute dimension of an action
$\frak{h}$. Then, relative dimensions (of the compo\-nents of
$\mathbf{p}$)
$[p_{\color{dark-green}0}]=[\frak{h}/{\color{dark-green}t}]=$ energy
and $[p_{\color{blue}a}]=[\frak{h}/{\color{blue}\ell}]=[m\,v]=$ 3d
momentum, q.e.d.. As we know from Lagrangian formalism, the action is
a 4d scalar. Accordingly, the absolute dimension of 4-momentum is
represented by the 4d scalar $\mathfrak h$ in SR and in GR.

We will proceed in a similar way in electrodynamics: We assume the
absolute dimension of the excitation $G$ to be that of an electric
charge $[G]=q$, for the field strength $F$ that of a magnetic flux
$[F]=\phi$. In SI, we have
$[G]\stackrel{\text{SI}}{=}\text{coulomb = C}$ and
$[F]\stackrel{\text{SI}}{=}\text{weber = W\hspace{-1pt}b}$. Thus,
besides the electric charge, introduced as dimension already in the
last section, we find a further fundamental physical dimension in
electrodynamics, namely 

\noindent$\bullet\,${\bf magnetic flux} $\phi$  (in SI $\rightarrow$
W\!b = V\hspace{1pt}s). Accordingly,
electric charge $q$ and magnetic flux $\phi$, both SR and GR scalars,
are the fundamental building blocks of modern electrodynamics. Thus,
{\it coulomb} and {\it weber} have a preferred status in SI.

But before proceeding to further consequences, we will check whether
these results coincide with our earlier knowledge from 3d. We read
off from \eqref{GF4d} that
\begin{align}\label{DH3d}[\bm{\mathcal{D}}]&=q,\quad\,[\bm{\mathcal{D}}_{ab}]=
  \frac{q}{\ell^2}\stackrel{\text{SI}}{=}\frac{{\rm C}}{{\rm m}^2}\,,\\
  [\bm{\mathcal{H}}]&=\frac{q}{t},\quad\,[\bm{\mathcal{H}}_a]=\frac{q}{t\ell}
       \stackrel{\text{SI}}{=}\frac{{\rm C}}{{\rm s\,m}}
       =\frac{{\rm A}}{{\rm m}}\,,
\end{align}
and 
\begin{align}\label{EB3d}[E]&=\frac{\Phi}{t},\quad\,
[E_a]=\frac{\Phi}{t\,\ell}\stackrel{\text{SI}}
{=}\frac{{\rm Wb}}{{\rm s\,m}}=\frac{{\rm V}}{{\rm m}},\\
[B]&=\Phi,\quad\,[B_{ab}]=\frac{\Phi}{\ell^2}\stackrel{\text{SI}}
     {=}\frac{{\rm Wb}}{{\rm m}^2}=\frac{{\rm V\,s}}{{\rm m}^2}={\rm T}
     \,,\quad\text{q.e.d.}
\end{align}
This proves that our attributions of $[G]=q$ and $[F]=\phi$ are correct.

An immediate consequence is that dimensionwise their product is an
action and their quotient an admittance:
\begin{align}
&[G]\!\times\![F]=q\,\phi={\mathfrak h}\stackrel{\rm
    SI}{=}{\rm A\,s\,Vs=V\,A\,s}^2={\rm J\,s}\,,\\
&{[{G}]}\,/\,
  {[{F}]}=\frac{q}{\phi}=\frac{q}{\frak{h}/q}=\frac{q^2}{\frak{h}}\stackrel{\rm
    SI}{=}\frac{{\rm A}^2{\rm s}^2}{{\rm V\,A\,s}^2}=\frac{{\rm A}}{
                          {\rm V}}=\frac{1}{\Omega}\,.
\end{align}

\section{Constitutive relations in electromagnetism: electric
  resistance as physical dimension}

In \eqref{13Maxwell} or in \eqref{Max4d}, we have the complete set of
the premetric Maxwell equations. They have to be supplemented by the
{\it constitutive relations} describing the medium/material under
consideration. Only then the emerging equations allow to predict the
temporal development of the medium. The special case of the vacuum is
described by \eqref{vacuum}.

Compared to vacuum, the next degree of complexity is a {\it local and
  linear} medium, which we will formulate directly in 4 dimensions. We
will use here literally the tensor calculus provided in Post's
authoritative book \cite{Post:1962}. There, the premetric Maxwell
equations read
$ \partial_\nu\mathfrak{G}^{\lambda\nu}={\mathfrak J}^\lambda$ and
$\partial_{[\kappa}F_{\lambda\nu]}=0$. The local and linear
constitutive law, relating excitation and field strength, is
represented by
\begin{equation}\label{constit1}
  \mathfrak{G}^{\lambda\nu}=\frac
  12\,\chi^{\lambda\nu\sigma\kappa}F_{\sigma\kappa}\,,\hspace{3.5pt}\text{with}
  \hspace{3.5pt}
  \chi^{\lambda\nu\sigma\kappa}=-\chi^{\lambda\nu\kappa\sigma}=
  -\chi^{\nu\lambda\sigma\kappa}\,,
\end{equation} 
where $\chi^{\lambda\nu\sigma\kappa}$ is a {\it constitutive tensor
  density} of rank 4 and weight $+1$, with the dimension
$[\chi]=[G]/[F]=1/{\rm resistance}$, with 36 independent components,
see \cite[p.31, Eq.(2.12)]{Post:1962}. Here $\chi$ can be decomposed
irreducibly under the linear group $GL(4,R)$ into the principal piece
(20 independent components), the skewon piece (15), and the axion
piece (1). Accordingly, the

\noindent $\bullet\,${\bf electric resistance} (in SI $\rightarrow$
$\Omega$) belongs to the fundamental dimensions in electrodynamics.

Let us consider the wave propagation in a local and linear medium
described by means of the tensor density in \eqref{constit1}. Then it
turns out that in 3d space in the geometric optics approximation the
waves span a (quartic) Kummer surface. These Kummer surfaces are
determined by the quartic algebraic equation
$\mathcal{K}^{\lambda\nu\sigma\kappa} k_\lambda k_\nu k_\sigma
k_\kappa=0$, with the wave covector $k_\mu$ and the Kummer tensor
density $\mathcal{K}^{\lambda\nu\sigma\kappa} (\chi)$, which itself is
defined as an expression cubic in terms of the constitutive tensor
density $\chi^{\lambda\nu\sigma\kappa}$; for details see Baekler et
al.\ \cite{Baekler:2014kha} and for applications Favaro et al.\
\cite{Favaro:2015jxa,Favaro:2016}. In vacuum, these Kummer surfaces
degenerate to light spheres, see \cite{Lammerzahl:2004ww}.

The special case of the vacuum is determined by
\begin{align}\label{vacuum1}
  \chi^{\lambda\nu\sigma\kappa}&\stackrel{\text{SI}}{=}\sqrt{
    \frac{\varepsilon_0}{\mu_0}}
  \sqrt{-g}\left(g^{\lambda\sigma}g^{\nu\kappa}-g^{\nu\sigma}
    g^{\lambda\kappa}\right)\,,\\ &\text{with}\quad\lambda_0
  :=\sqrt{\frac{\varepsilon_0}{\mu_0}}=:\frac{1}{\Omega_0}
  \approx \frac{1}{377\>\Omega}\,,
\end{align}
with the electric constant $\varepsilon_0$ and the magnetic constant $\mu_0$,
see Post \cite[Eqs.(9.4) with (9.18)]{Post:1962}. Incidentally, for
the speed of light, we find $c=1/\sqrt{\ve_0 \mu_0}$. In exterior
calculus, the vacuum excitation and the field strength are related as
follows: $G=\lambda_0\,^\star\!F$, with the metric dependent Hodge
star operator $^{\star}\,$.

We hope that it became clear that the premetric Maxwell equations in
Eqs.\eqref{13Maxwell} or \eqref{Max4d} do {\it not} contain the
metric tensor. The latter only enters the constitutive law in
Eq.\eqref{vacuum} or in Eq.\eqref{constit1} together with the
constitutive tensor density \eqref{vacuum1}.

\section{SR-scalars and GR-scalars, emergence of quantum 
aspects: elementary charge and fluxon}

Fleischmann \cite{Fleischmann:1971} observed, in a seemingly widely
overlooked paper, that in physics we have, on the one side, 4d laws
which are not changed by {\it affine transformations} and, on the
other side, those 4d laws in which the metric tensor enters. We just
had a perfect example: The premetric Maxwell equations belong to the
former class, the constitutive relations to the latter class.

Fleischmann specializes these considerations also to scalar
quantities.  What he calls ``metric invariant scalars,'' are scalars,
which do {\it not} depend on the metric and are diffeomorphism
invariant. We call them (metric-indepen\-dent) {\sl GR-scalars.} With
Lorentz scalars he designates scalars that do depend on the metric and
are only invariant under Lorentz transformations; we call them
SR-scalars. From the context it is clear that Fleisch\-mann really
considers inhomogeneous Lorentz transformations, also known as
Poincar\'e transformations. The Poincar\'e group is the group of motions of
flat Minkowski space, that is, when gravity can be neglected.

Since universal constants are assumed to be scalars, we can divide
them into GR-scalars and SR-scalars. As we already saw earlier in our
paper, the electric charge $q$, the magnetic flux $\phi$, the action
$\mathfrak h$ are apparently GR-scalars, and the same is valid for the
entropy $S$. And products and quotients of GR-scalars are again
GR-scalars. Since $[q]\times [\phi]=[\mathfrak h]$, fundamental
GR-scalars are expected to be expressible as follows:
\begin{equation}\label{GR-scalars}
{q^{n_1} \mathfrak{h}^{n_2}=\;{\rm 4d\; scalars}}\qquad
(n_1, n_2\;{\rm numbers})\,.
\end{equation}
Specifically, we observe
 \begin{equation}\label{xxx}{
    q}\rightarrow {\rm el.\ charge},\; \frac{{
      \mathfrak{h}}}{{
      q}} \rightarrow {\rm mg.\ flux},\;\frac{{
      \mathfrak{h}}}{{
      q^2}} \rightarrow {\rm el.\ resistance},...
\end{equation}
We find phenomenologically only $n_1=\pm1,-2;n_2=0,1$. In this way,
the fundamental quantities in classical electrodynamics are exhausted.

The speed of light $c$ is only a SR-scalar, since light is
influenced by gravitation in a direct way: starlight gets deflected by
the gravitational field of the Sun, for example. Incidentally, the
speed of light $c$ is the only SR-scalar amongst the universal constants
that is known to us.

In the theory of dimensions, we considered in Sec.5 the set
$(\mathfrak{h}, q, \ell, t)$, in SI $(m, I, \ell, t)$,
{with $t$ as defined via a spectral line of $^{133}$Cs and
  $\ell$ via the fixed speed of light}.  Now, after identifying charge
and magnetic flux as fundamental quantities, we can choose
alternatively $(q,\Phi, \ell, t)\stackrel{\rm SI}{=}$(C, Wb, m, s) or
$(q, \mathfrak{h}, \ell, t)\stackrel{\rm SI}{=}$(C, J s, m, s).

Up to now{---apart from the SI-definition of the
  second---}in our whole paper we did not address any quantum aspects
explicitly. {{\it All our considerations were in the
    framework of classical mechanics and classical electrodynamics.}}
But now, suddenly we recognize that some of the GR-scalars mentioned
can display quantum aspects. In nature, interestingly enough, all of
those GR-scalars are related to quantum effects.

In nature, the electric charge is quantized and only occurs in the
form of elementary electric charges (or in quarks as
$\pm \frac{1}{3}e,\pm\frac{2}{3}e$):
\begin{equation}\label{elementarycharge}
e\stackrel{\rm SI}{\approx}1.6 \times 10^{-19}\,
    {\rm C}\qquad \text{(GR-scalar)}\,.
\end{equation}
The magnetic flux exists in an unquantized form. However, under
suitable conditions, in a superconductor of type II, we observe a flux
line lattice composed of single flux lines (Abrikosov vortices) each
of which carries one magnetic flux unit, a quantized fluxon of
\begin{equation}\label{fluxon}
  \Phi_0=h/(2e)\stackrel{\rm SI}{\approx} 2.1\times
  10^{-15}\,{\rm Wb}\qquad \text{(GR-scalar)\,,}
\end{equation}
see Figure 2. The factor $2$ appears in \eqref{fluxon} since the
superconductivity is induced by the Cooper pairs, which carry two
unit charges.

\begin{figure*}
\hspace{-70pt}
\includegraphics[width=17truecm]{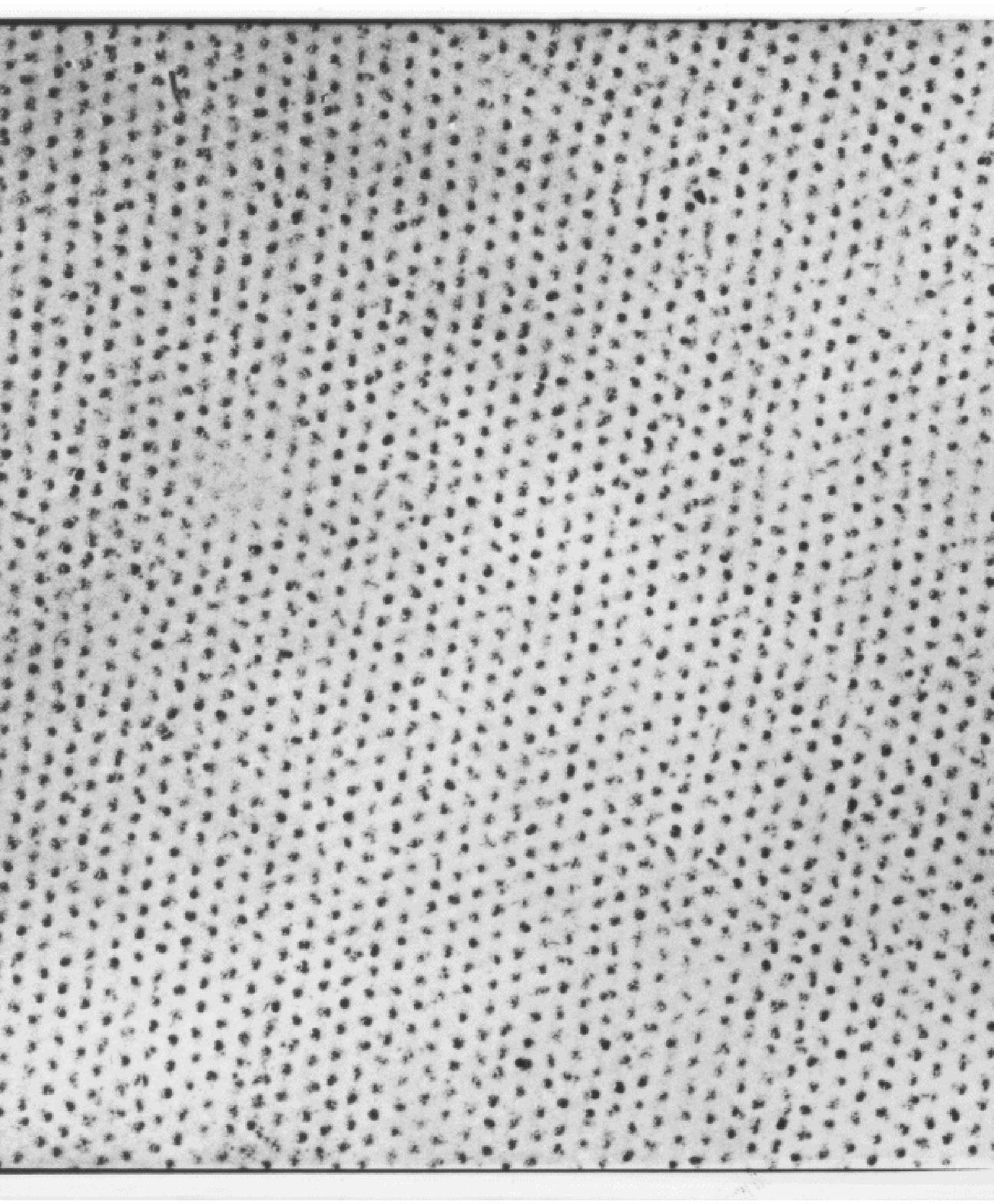}\vspace{-40pt}
\caption{\label{Fig.2}{Flux lines in type II superconductors according
    to Essmann \& Tr\"auble \cite{Essmann:1967} (original by courtesy
    of U.\hspace{1pt}Ess\-mann): Niobium disc (diameter 4 mm, thickness 1 mm), 1.2
    K, ${\cal H}=78\,{\rm kA/m}$. Parameter of flux line lattice 170
    nm.}}
\end{figure*}

\section{Josephson constant $K_{\text{J}}$ and von 
Klitzing constant $R_{\text{K}}$ as GR-scalars, Kibble 
(or watt) balance}
  
If we pick in \eqref{xxx} for $q$ the elementary charge $e$ and
for $\frak{h}$ the Planck constant $h$, then we arrive at the {\it
  Josephson\/} and the {\it von Klitzing\/} constants of modern
metrology (peta=P=$10^{15}$):
\begin{align}\label{JvK}
K_{\rm J} &= {\frac {2e} {h}}\stackrel{\rm SI}{\approx} 0.483 \;{\rm
  \frac{PHz}{V}} \approx \frac{1}{2.068 \times 10^{-15} {\rm W\!b}},
\\  R_{{\rm K}} &= \frac{h}{e^2}\stackrel{\rm SI}{\approx} 25.813\;
{\rm k}\Omega\,.
\end{align}
{Since these constants can be measured (in the context of
  the old SI) or realized (within the new SI) with very high
  precision, they also give very precise measurements or realizations
  of $e$ and $h$. More specifically, in the new SI, the elementary
  charge $e$ and the Planck constant $h$ have fixed values. This
  implies that the same will be true for $K_{\rm J}$ and $R_{\rm K}$
  as well, see G\"obel and Siegner \cite[p.122]{Goebel:2015}. The new SI of
  post-2018, will be built on
  the GR-scalars $K_{\text{J}}$ and $R_{\text{K}}$, since the
  Josephson and the von Klitzing (QHE) effects belong to the most
  precise tools in metrology.} The {\it M\"ossbauer effect} is of
similar precision as the Josephson
and the quantum Hall effects. However, it measures frequencies or
energies which are not GR-scalars. Thus, the M\"ossbauer effect turns
out not to be useful for fundamental
metrology.\footnote{{Still, for the definition of the
    second a frequency is used as a fundamental entity. We have then
    to refer it always to the rest system in which the frequency is produced.}}

{We know that {$K_{\text{J}}$ and $R_{\text{K}}$ are true GR-scalars},
  since they have the dimension of a reciprocal magnetic flux and a
  resistance, respectively.} By the same token, the Quantum Hall
Effect is not influenced by the gravitational field,\footnote{A
  somewhat related question was investigated by Russer
  \cite{Russer:1983}: He computed the effect of an {\it acceleration}
  on a Josephson junction.} as discussed by Obukhov, Rosenow, and Hehl
\cite{Hehl:2003rs}.

For substituting the kilogram prototype by a new definition for the
kilogram, we need the Kibble balance. We will discuss it shortly:
  
\subsubsection*{Quantum definition of the kilogram -- the Kibble (or
  watt) balance}

{The most fundamental physical unit is the `second,' the unit
  of time. The second it defined as the duration of a certain number
  of periods of the radiation related to a certain atomic
  transition. For doing so, the apparatus and the atom have to be at
  the same position in the same frame of reference. These periods, as
  measured in other frames, are related by the laws of Special and
  General Relativity, or some generalizations of it. The definition of
  the second is the same independently where and when and in which
  frame of reference this definition is made. This definition is
  independent from any particular physical law or symmetry or geometry
  of spacetime. Only the relation to other reference frames or
  positions rely on physical laws and symmetries.

  Accordingly, we can take the definition of the second for
  granted. Then, in SI, there is a numerical value assigned to the
  speed of light, namely c = 299 792 458 ms$^{-1}$. In turn, the meter
  can be obtained in terms of the definition of the second; it is no
  longer an independent unit.

  For the definition of the third mechanical unit, the
  \textit{kilogram}, a constant is needed which contains, besides the
  units of time and length, the unit of mass. One option for that is
  the Planck constant $h$. It is experimentally very difficult to
  directly relate a mechanical unit with a constant that is
  characteristic for the quantum regime of physics. The proposed
  procedure is to first relate the mass to electromagnetic quantities
  and, subsequently, to link these electromagnetic quantities to those
  of quantum physics. One experimental device which relates the Planck
  constant to a mass is the Kibble or watt balance.

  This balance uses the force produced by a current-carrying wire in a
  magnetic field to balance the weight of a mass. By taking
  measurements of other experimentally-derived quantities and by using
  a given measure for the kilogram, the Kibble balance can be used to
  accurately measure the Planck constant $h$. The idea now is to take the
  Planck constant $h$ and to relate that to a mass unit.}

Establishing the relation between the mass and electric unity needs
two independent experiments carried through by the same apparatus:
\begin{itemize}
\item 1st experiment: We have an electric coil of a certain length $L$ 
  which moves through a magnetic field $B$ which is pointing
  outwards. Then the induced voltage $U$ relates to the velocity $v$
  according to
\begin{equation}
	U = B L v\,.
\end{equation}
\item 2nd experiment: the gravitational force acting on a mass $m$ is
  $m g$. If this mass is attached to the coil, then the gravitational
  force can be counterbalanced by magnetic levitation force $F$
  induced by a current $I$ through the coil, $F = I B L$. Equating
  both forces then yields $m g = I L B$. Combining both experiments
  eliminates $BL$ and yields
\begin{equation}
	m g v = U I \, ,      
      \end{equation}
    where on the left hand side we have mechanical, and on the right
      hand side electric quantities.
    \end{itemize}
    Now, the next step is to determine the voltage $U$ and the current
    $I$. This is where the quantum effects, namely the Josephson
    effect and the quantum Hall effect come in and provide the
    relation to the Planck constant.

A superconducting material is divided into two domains by means of a
thin insulator, which acts as potential wall. The Cooper pairs in the
superconducting regions are still coupled with each other due to the
tunnel effect. The AC Josephson effect relates the voltage $U$ between
these two regions to the frequency $\nu$ of the current through the
insulator
\begin{equation}
\nu = 2 \frac{e}{h} U\,,
\end{equation}
where $e$ is the elementary electric charge of the electron.

The quantum Hall effect is a quantum effect of electrons subject to a
magnetic field in a restricted geometry and yields a unit of
electrical resistance
\begin{equation}
R_{\text{K}} = \frac{h}{e^2} \, ,
\end{equation}  
called the von Klitzing constant, so that each resistance is a
multiple of this unit: $R = i R_{\text{K}}$ where $i$ is a positive
integer.

With this we are able to relate the mass $m$ to the Planck
constant. In
\begin{equation}\label{mystery}
m = \frac{1}{g v} U I\,,
\end{equation}
we replace $I = U^\prime/R^\prime$, with $R^\prime = i R_{\text{K}}$.
In \eqref{mystery} we have on the right hand side $U$ and $I$. They
are measured in two different experiments with the same apparatus, as
indicated before. With the Josephson effect, we can compare a voltage
$U$ with a frequency according to
{$U =  \frac{h}{2e} \nu$.} Furthermore, we
can realize a current with another voltage $U^\prime$ and a resistance
$R_{\rm K}$. Both results can be expressed in terms of $e, h$,
and $\nu$. All of this we substitute into \eqref{mystery} and we find
the following result:
\begin{equation}
m = \frac{i}{4} \frac{\nu \nu^\prime}{g v} h \, . \label{mhkilo}
\end{equation}
Here, $\nu$, $\nu^\prime$, $v$, and $g$ are based on the second and
the meter, which itself is based on the second. Therefore, by defining
$h$, we have---via the Kibble balance---an experimental realization of
the kilogram. For a more detailed description of the Kibble balance
and its technical realization, one may compare {Stock
  \cite{Stock},} Steiner \cite{Steiner:2013}, and Robinson et al.\
\cite{Robinson:2016}.  Incidentally, one may also solve \eqref{mhkilo}
for $h$ and then interpret this as a measurement of $h$, based on the
kilogram represented the the prototype.

Note that only the {\it gravitational} mass enters our description of
the Kibble balance. The {\it inertial} mass does not play a
role. However, according to a suggestion of Bord\'e \cite{Borde:2018},
one may redo this experiment in outer space and replace the
gravitational force by the centrifugal force. Then only the inertial
mass enters this procedure.  It is surprising that with these
procedure it seems to be possible to independently measure the
inertial and the gravitational mass. Usually one only can determine
the ratio of both masses which originates from the Newton axiom
${\bf F} = m_{\text{i}} {\bf a}$ with
${\bf F} = m_{\text g} {\boldsymbol\nabla} U$, where $U$ is the
Newtonian gravitational potential.

\begin{footnotesize}
  {If we distinguish between inertial and gravitational mass,
  then---owing to the two possible measurements of $h$ using the watt
  balance---one may also speculate on the distinction between an
  inertial and a gravitational Planck constant. In fact, in the
  Schr\"odinger equation only the ratio of $m/\hbar$ appears in the
  kinetic as well as in the gravitational interaction term. Then the
  corresponding ratios may be introduced as
  $m_{\text{i}}/\hbar_{\text{i}}$ and
  $m_{\text{g}}/\hbar_{\text{g}}$. Equivalently, in the conventional
  Schr\"odinger equation the $\hbar$ entering the time derivative may
  be different from the $\hbar$ appearing in the kinetic term. Thus,
  the distinction between inertial and gravitational mass opens up the
  question of a distinction between different Planck constants, see,
  however, \cite{Fischbach:1991eg} and \cite{Petley:1992pk}. This
  hints at a possible deep connection between gravitation and quantum
  theory---but a much deeper analysis is required in this
  context.}\end{footnotesize}

Now, after having defined the unit of mass, we can derive the unit of
energy, the joule.

\subsubsection*{{Completion of the SI}}

{We concentrated in this essay on mechanics and on
  electromagnetics. However, the complete new SI also encompasses
  notions of thermodynamics and chemistry. Namely the entropy $S$ and
  the Boltzmann constant $k_{\rm B}$, with $[k_{\rm B}]=$
  energy/temperature $\stackrel{{\rm SI}}{=}$ joule/kelvin = J/K;
  moreover, the Avogadro constant $N_{\text{A}}$ as the number of
  atoms/molecules in one mole, which is independent from the previous
  units.\footnote{We suppressed here the luminous efficacy since it
    is not so important for fundamental questions.}  Whereas within
  the SI the Boltzmann constant $k_{\rm B}$ just plays the role of a
  mere conversion factor, its deeper physical meaning lies in the
  introduction of the statistical entropy.

  Accordingly, besides the
  frequency normal of the Caesium atom $\Delta f_{\rm Cs}$ of the
  standard atomic clock, the fundamental constants in the new SI are,
  $h, e, k_{\rm B}, c, N_{\text{A}}$. For a comprehensive account we
  refer to \cite{Goebel:2015}}.


\section{Can we measure or define the speed of light in a
  gravitational field?}

``{\it When you shoot a ray of light parallel to the black hole, the
  local speed of light is less than the speed of light at
  infinity.}''\footnote{Oral statement by Juan Maldacena, discussion
  session during PITP 2018 at the Institute for Advanced Study,
  Princeton, 26 July 2018 https://pitp.ias.edu/program-schedule-2018
  (Jens Boos, private communication). {We leave the
    judgment on this sweeping statement to the discretion of our
    readers. The analogous applies to the quotation of a string
    theoretician in Sec.1.}} \hfill J.~Maldacena (2018)\bigskip

We can distinguish two aspects of the speed of light in a
  gravitational field: 
\begin{itemize}
	\item the speed of light in vacuum is unique,
	\item the speed of light depends on the gravitational field. 
\end{itemize}

\noindent Within standard physics, that is, the standard Maxwell
equations in vacuum and GR, there is only one light ray in a given
direction (assuming that the environment is small enough---this
statement does not hold if the environment considered incorporates a
Black Hole). This is taken as an axiom in the constructive axiomatic
scheme of Ehlers, Pirani, and Schild \cite{Ehlers:1972} which yields
an axiomatic foundation for the Riemannian geometric structure of
GR. This uniqueness also holds true in the tangent space at each point
in spacetime, which is assumed to be a smooth manifold. Within the
framework of a premetric Maxwell theory with local and linear
constitutive law, this uniqueness is only valid if there is {\it no
  birefringence} \cite{Lammerzahl:2004ww}, what has been
experimentally proven with extremely high accuracy, see Kostelecky,
Mewes \cite{KosteleckyMewes02} and Ni \cite{Ni:2015rdf,Ni:2017ika}. It
has also been shown that the maximum velocity of massive particles
coincides with the unique speed of light, see e.g.\
\cite{LaemmerzahldeVirgilio2016}.

The next step is the following: We take a point $P$ of spacetime and
consider all light rays starting at $P$. After a small time interval
$dt$, they generate a 2d-surface. If the spacetime geometry is Riemannian,
then we can find a coordinate system such that the 2d-surface is a
sphere.

In more general geometries, like in Finsler geometry, no such
coordinate system can be found. Nevertheless, the surface of the light
rays after a time span $dt$ can be taken to {\it define} a unit
sphere. Then the speed of light has, by definition, some given, fixed
value. This is the procedure presently used whereat the speed of light
has the fixed value of 299 792 458 m/s. Given the unit of a second of
a certain transition of the Cs atom, this uniquely defines the
meter. This procedure holds in any geometry, even in a Finsler
geometry \cite{LaemmerzahlPerlick2018}. Consequently, the uniqueness
of defining a speed of light only requires the vanishing of
birefringence.

In a recent paper of Braun, Schneiter, and Fischer
\cite{Braun:2015twa}, the precision was discussed with which the speed
of light can be measured. This can be of relevance in the case that, e.g.,
spacetime fluctuations of quantum gravitational origin (or from
perturbations of the spacetime metric from the signal itself) yield
fundamental limits in the precision of a signal transfer. In turn,
this will give a fundamental minimum precision in the definition of
the meter. However, until now this is far from experimental reach.

If the vacuum for some reason (maybe due to not yet known effects from
a not yet fully worked out theory of quantum gravity) turns out to act
birefringent, then the present SI system will break down and new
concepts have to be found. Since the speed of light is not only the
speed of light but also the maximum speed of massive particles, one
may replace the speed of light by one mathematically particularly
nice maximum speed. In such a case, the precision of the definition
of the meter certainly will be much worse than the present one. It
probably needs more theoretically investigations of what kind of
concepts should be taken to replace the old definition. Another aspect
is that taking into account that the speed of light is a dynamic
phenomenon relying on a certain physical theory, namely Maxwell theory,
whether it is possible to replace this concept by something which is
independent of physical theories.

A further aspect in this context is the speed of gravity, see, for
instance, the discussion of Unnikrishnan and Gillies
\cite{Unnikrishnan:2018}. The speed of light as well as the maximum
speed of particles all take place in the background of a certain
spacetime geometry. Now, current theories of gravity predict
gravitational waves which have been confirmed in various ways. The
speed \textit{of} gravity, in particular of gravitational waves, is
completely different from any speed of objects \textit{within} the
gravitational field. Even if the speed of gravity has a value much
different from the speed of light, no fundamental principles related
to the dynamics of photons and other particles will be violated, see
also Ellis and Uzan \cite{Ellis:2003pw}. However, also this has to be
much more deeply analyzed, in particular in view of the fact that with
gravitational waves one can transport information and also energy.

Recently the event GW170817 was observed gravitationally and
electromagnetically likewise. Perhaps the delay time between these
measurements elucidates the problem whether $c_0$ is really a true
GR-scalar, see, e.g., Wei et al.\ \cite{Wei:2017} and Shoemaker et
al.\ \cite{Shoemaker:2017}. However, any delay between the
electromagnetic and gravitational signals could also be due to a
slight delay due to their production, about which there could
evidently be some uncertainty.

The above approach, using a constant speed of light, relies on the
spacetime being a smooth differentiable manifold which allows to
define tangent spaces. The definition of the SI units, in particular
within the frame of the new system, is taking place in this tangent
space. In this sense the speed of light is a GR scalar.

This has to be distinguished from the observation that---taking a
global view beyond the tangent space---the speed of light depends on
the gravitational field. The most prominent experimental fact related
to that is the gravitational time delay confirmed with a $10^{-5}$
precision by the Cassini experiment. The time a signal needs from the
Cassini satellite to reach the Earth becomes larger if the signal
travels through a gravitational field. In this sense the {\it
  gravitational field} acts like a {\it refractive index.} And this
clearly demonstrates that the speed of light in a gravitational field
in the context of a nonlocal view does not coincide with its vacuum
value: $c=c(g)$. As a consequence, in harmony with the conclusions of
Fleisch\-mann \cite{Fleischmann:1971}, this speed of light $c$ is only a
SR-scalar---in contrast to $h,e,...$. It belongs to the Fleischmann
class 2.

The latter result can be obtained also by means of the first approach:
one integrates the coordinate time of the light ray along its
trajectory. This can be transformed to an observer's local proper time
by another transformation given by the metric component
$g_{tt}$. Therefore, the first approach is consistent with the second
one provided light rays do not show birefringence. It also shows that
the defined speed of light is a scalar. This is because the defined
value of the speed of light is independent of any direction. It is the
speed of all light rays emanating from a spacetime point.

\section{{Closing statement}}

The new definition of the system of physical units, decided in November
2018, is a major step towards clarity and uniqueness of SI. The new
definition solely rests on the definition of natural constants, whereas
the corresponding experimental realization is left open. This gives
natural space for the introduction of new experimental methods.

The new system is much clearer on the theoretical side than the old
system. However, there are still some issues which needs a better
analysis. One point is the interplay between laws of nature and the
(wo)man-made definitions or conventions. It still has to be analyzed whether
the introduction of physical units can be carried through without any
reference to physical laws like the Maxwell equations or GR.

The new systems relies more on quantum mechanics than the old
one. This has many advantages, see \cite{Lammerzahl:2018}, e.g.: (i)
quantum mechanics and, thus, quantum systems are unique and do not
require the precise manufacturing of, e.g., prototypes. Atoms are the
same everywhere in the universe. (ii) Quantum mechanics also gives a
clear definition of quantum systems by means of a finite number of
rational numbers. This enables an easy dissemination of units just by
agreeing on such numbers. Dissemination through the transport of
prototypes is no longer necessary. (iii) On the technical side, the
use of quantum systems has a huge potential of miniaturization.

A further point mentioned in this article and still to be discussed is
the different nature of the various fundamental constants used, for
the speed of light, see Peres \cite{Peres:2002rj}, and for the speed
of gravity, Unnikrishnan and Gillies \cite{Unnikrishnan:2018}. The
speed of light is a constants which is derived from a propagation
phenomenon. The electric charge is a coupling constant. The Planck
constant is a conversion factor between energy and frequency, but is
also fundamental in the uncertainty relation. However, effectively
only the ratio between the Planck constant and mass appears in the
Schr\"odinger equation as well as in the uncertainty relation. The
Boltzmann constant is a conversion factor, too, conversion between
temperature and energy, but also required in the relation between
statistical entropy and thermodynamical entropy. (The examples where
these fundamental constants appear is not meant to be complete.) These
examples show that the nature of these constants is very different and
that it is not clear which aspect of these constants is primary in the
definitions for the SI.

In summary, the new SI system is a big step in the methodological
unification and simplification of the introduction of physical
units. It still requires, however, a better understanding of the
nature of the constants used; also which realization of which constant
is really required for establishing the SI. Moreover, analysis is
desirable of the dependencies of all definitions used on the physical
laws of nature, including the symmetries like the Einstein equivalence
principle.

\subsubsection*{Acknowledgments}

Both authors would like to thank the organizers of the Wilhelm and
Else Heraeus workshop for the invitation: to Klaus Blaum (Heidelberg),
to Dmitry Budker (Mainz), and to Andrey Surzhykov
(Braunschweig). Moreover, FWH is most grateful to Yuri Obukhov
(Moscow) and Yakov Itin (Jerusalem) for past collaborations on the
issue of physical dimensions. We also acknowledge very helpful remarks
by Jens Boos (Edmonton), Alberto Favaro (London), Michael Krystek
(Berlin), James Nester (Chung-li), and Volker Perlick (Bremen). And
last, but not least, we thank Joachim Ullrich (Braunschweig) for
useful hints and interesting discussions and two anonymous referees
  for their careful evaluations. CL acknowledges support of the DFG
  funded Research Training Group 1620 ``Models of Gravity'' and of the
  Collaborative Research Center (SFB) 1128 ``geo-Q'' and 1227
  ``DQ-mat.''



\end{document}